\documentstyle[prl,aps,multicol]{revtex}
\def\q{\mbox{\boldmath $q$}}

\input{epsf}

\begin{document}
\setcounter{page}{1}
\draft

\title{{\large
\bf Chaotic dynamics of a classical radiant cavity}
}

\author{Giuliano Benenti$^{(a,b,c,*)}$, 
Giulio Casati$^{(a,b,c)}$ and 
Italo Guarneri$^{(a,b,d)}$}

\address{$^{(a)}$International Centre for the Study of Dynamical Systems,}
\address{
Universit\`a di Milano, sede di Como, via Lucini 3, 22100 
Como, Italy}
\address{$^{(b)}$Istituto Nazionale di Fisica della Materia,
Unit\`a di Milano, Via Celoria 16, 20133 Milano, Italy}
\address{$^{(c)}$ INFN, Sezione di Milano,  Via Celoria 16, 20133 Milano, Italy}
\address{$^{(d)}$ INFN, sezione di pavia, Via Bassi 6, 27100 Pavia, Italy}

\date{October 5, 1998}
\maketitle

\begin{abstract} 
The statistical properties of a classical electromagnetic field in 
interaction with matter are numerically investigated on a one--dimensional 
model of a radiant cavity, conservative and with finite total energy. 
Our results suggest 
a trend towards equipartition of 
energy, with the 
relaxation times of the normal modes of the cavity increasing with 
the mode frequency according to a law, the form of which depends on the 
shape of the charge distribution. 
\end{abstract} 

\begin{multicols}{2}
\narrowtext

The study of quantum deviations from classical 
ergodicity has occupied much of Quantum Chaology since its origins.  
Remarkably enough, the historical development of quantum mechanics 
started with the Blackbody problem, which displays a deviation as blatant as 
possible 
from classical ergodicity. When 
a classical radiation field interacts with 
matter inside an enclosure with perfectly reflecting walls, approach to 
statistical equilibrium -- if at all possible -- appears to entail 
unending escape of energy towards higher and higher frequencies, in sharp 
contrast to the Planck distribution law\cite{JEANS}. Thus the problem of 
blackbody radiation was at once the first problem in quantum mechanics, 
the first problem in quantum field theory, and the first problem in 
quantum chaos.
Considerable progress has been meanwhile attained in understanding the 
complex behaviour of nonlinear classical dynamical systems and its 
quantum counterparts, so  a re--examination of the blackbody problem in 
the light of such developments appears necessary. One would like, 
first, to  
build a Hamiltonian model of a radiant cavity, which does indeed exhibit  
the sort of tendency to equipartition expected in Jeans's time; second, 
to understand how does Planck's law emerge from the quantal dynamics 
of that very model. Neither of these issues seems to have been satisfactorily 
dealt with as yet. In this Letter we accomplish the first part of the task 
by presenting a model whose classical dynamics leads to equipartition in the 
sense of Jeans.
\newline\indent
Our model is a variant of one which was  
introduced years ago \cite{BCL72} to this purpose, and which was later  
investigated in several papers 
\cite{BLVG74}.
None of those investigations was able to detect a 
tendency towards energy equipartition
among the normal modes of the cavity. 
Analogous results were obtained on different 
models \cite{GAFGG89,P83}, such as a one--dimensional linear 
string interacting with nonlinear oscillators etc.
\newline\indent
The general picture which emerges from all these works, in which
the Newton--Maxwell equations were numerically solved, is that there is 
no tendency to energy equipartition among the field normal modes.
The reason lies with two fully 
general aspects of the classical field--matter interaction. First, 
the total energy is finite, whereas the number of freedoms is infinite; 
second, field modes can only exchange energy via interaction with finitely 
many mechanical freedoms. Thus  mechanical 
nonlinearities  
become less and less effective  as 
energy flows from the matter to the field: this fact
prevents the appearance of an altogether chaotic dynamics, thus causing 
high frequency modes to be nonergodically ``frozen''.
One therefore needs a model, giving rise to 
chaotic behavior of the mechanical freedoms, 
no matter how small their energy is.  
\newline\indent
Though extremely simplified, our model displays 
this property. Let us first consider   
\cite{BCL72} an electromagnetic field confined inbetween two 
parallel, perfectly reflecting plane mirrors, a distance $2l$ apart.
We take Cartesian coordinates $XYZ$  
with the $X$ axis normal to the mirrors, and restrict 
to excitations only 
dependent on $X$, thus getting a 1--dimensional radiant cavity, the normal 
modes of which have angular frequencies $\omega_n=(\pi c/2l)n$, $n=1,2,...$.
Then we introduce a uniformly charged, infinite plate of thickness $2\delta$,
situated midway 
between the mirrors and parallel to them, bound to move along the 
$Z$--direction only. We denote $z$ its displacement in that direction, 
$\sigma$ and $m$ the charge and mass densities per unit surface of the plate, 
$f(x)$ the normalized (transverse) distribution of charge in the plate. 
Finally, the plate is subject to a mechanical 
restoring force per unit surface, 
$F(z)=-m\omega_0^2z$.
Using the Coulomb gauge, plus zero boundary conditions on the mirrors 
for the $Z$ component 
of the vector potential, we obtain the following Hamiltonian 
for the full system plate plus field: 
\begin{eqnarray}
\begin{array}{c}
\displaystyle{
H_0=\frac{1}{2m}\left(p_z-2\left(\frac{\pi}{l}\right)^{1/2}
\!\sigma\sum_{n=1}^{\infty}{}^{'}a_n q_n\right)^2+}
\cr\cr
\displaystyle{
\frac{1}{2}m\omega_0^2 z^2+
\frac{1}{2}\sum_{n=1}^{\infty}{}^{'}(p_{n}^2+
\omega_n^2 q_n^2)},
\end{array}
\end{eqnarray}
where $(z,p_z)$ and $(q_n,p_n)$ are canonical conjugated variables for the 
plate and the $n-$th mode of the field respectively. In particular, 
$q_n(t)$ is the amplitude of the 
$Z$ component of the vector potential on the $n$--th normal mode 
of the free field, and 
the coefficients $a_n$ are given by
$a_n=\int_{-\delta}^{\delta}dx f(x)\cos(\omega_n x/c)$. $\sum$' means the 
sum over odd $n$'s only, because, with the chosen boundary conditions, 
even modes do not interact with the 
plate.

Finite--energy states of  
our  hamiltonian system correspond to vectors in the Hilbert space ${\cal H}_0$
of 
square--summable,  
$\infty$--dimensional vectors
\begin{equation}
\q=\left\{m^{1/2}\omega_0 z,m^{1/2}\dot{z},
...,\omega_{2n-1}q_{2n-1},
p_{2n-1},...\right\}
\end{equation}
whose squared norm is just twice the energy. Since Hamilton's 
equations are linear, the evolution of states in ${\cal H}_0$ is given     
by a unitary group $\exp(iH_0t)$, with a 
complete set of orthonormal eigenvectors , 
\begin{equation}
{\bf u}_{k}=C_k\left\{\frac{\omega_0}{\Omega_k},1,...,
\frac{\epsilon\,a_{2n-1}\,\omega_{2n-1}}
{\Omega_k^2+\omega_{2n-1}^2}\,,
\frac{\epsilon\,a_{2n-1}\,\Omega_k}
{\Omega_k^2+\omega_{2n-1}^2},...\right\},
\end{equation}
where $\epsilon=2\sigma(\pi/ml)^{1/2}$ and $C_k$ is a normalization 
constant. These eigenvectors define normal modes of the total system, with 
eigenfrequencies $\Omega_k$  
given by the 
imaginary roots of the secular equation:
\begin{equation}
\Omega_k^2\,\left(1+\epsilon^2\sum_{n=1}^{\infty}\,\frac{a_{2n-1}^2}
{\Omega_k^2+\omega_{2n-1}^2}\right)+\omega_0^2=0.
\end{equation}
        
We shall now introduce a nonlinear mechanism, which will couple these normal 
modes, giving rise to energy exchanges between them. To this end we
introduce a second plate,
parallel to the first, bound to move in the same direction, and with
the same mass density. We assume for simplicity this plate to carry 
no charge, 
so that its motion is not influenced by the field. The 
interaction between plates is purely mechanical, and simulated by 
elastic bounces when $|z(t)-z_1(t)|=R$, with $z_1$ the $Z$--coordinate 
of the neutral plate.

Between collisions, the system is integrable: the motion of the charged 
plate and the field is given by the action of the group
$\exp(iH_0t)$, while 
the second plate is moving at a uniform speed. This is most easily 
described by adding to the vector ${\bf q}$ 
one more component $m^{1/2}w$, with $w$ the velocity of the uncharged plate. 
The new Hilbert space ${\cal H}$ 
of such ${\bf q}$--vectors is the phase space of our model.    
The evolution between 
collisions 
is again unitary, given by $\exp(iHt)$, with a generator 
$H$, and a complete set  
of normal modes, which are trivially related to the above described ones.
At collisions the two plates exchange their velocities, 
thus mixing  all the amplitudes in the expansion of the state vector over 
normal modes.

The evolution from immediately after one collision to immediately after 
the next is given by a map, which, in Hilbert space notations, has 
the following simple form:
\begin{equation}
{\cal S}({\bf q}) =(Id-2P)e^{iHt({\bf q})}{\bf q},
\label{map}
\end{equation}
The 1st (operator) factor describes a collision:  
$Id$ is the identity operator, $P$ is a one--dimensional 
projection: $P{\bf q}=<{\bf e}\vert{\bf q}>{\bf e}$, where ${\bf e}$ is the 
unit vector such that the scalar product $<{\bf e}\vert{\bf q}>$ yields the 
relative velocity of the two plates in the state ${\bf q}$. 
The 2nd factor describes  evolution over the free-flight time  $t({\bf q})$, 
which is the smallest 
positive root of the equations
\begin{equation}
\vert z(e^{iHt}{\bf q})-z({\bf q})-w({\bf q})t\vert=0,2R.
\end{equation}
The map allows for efficient numerical simulation:  e.g., in the
case of $200$ oscillators, we were able to follow a trajectory up to  
 $5\times 10^6$ bounces with a relative error in energy conservation 
less than $10^{-10}$.

In numerical simulations, one has of course to consider a finite  number $N$ 
of field oscillators. In our computations we have varied $N$ and all   
other parameters, except 
$l=\pi,m=1,c=1$; moreover, since  the dynamics depends on
energy only via the scaled parameter $R/\sqrt{E}$, we have always taken $E=1$ 
and varied the ``free path'' $R$ instead.

The choice of the charge density $f(x)$ is important, 
because the coupling of individual modes to the charged plate is scaled 
by the coefficients $a_n$ of the Fourier expansion of $f(x)$. Choosing 
a singular density $f(x)$, as in earlier studies, results, at all times, in a 
power law decay of the distribution of energy over the field modes, 
so that 
truncation effects are already significant at small integration times. 
We have therefore chosen 
$f(x)=k\exp(-\delta^2/(\delta^2-x^2))$ (the standard 
compactly supported $C^{\infty}$ function), 
with the constant $k$ fixed 
from normalization; this ensures a 
faster than algebraic, albeit nonexponential, decay of the distributions.

Even though we do not have rigorous results, the dynamics
of this billiard--type model appears to be completely chaotic independently
of the total energy.
Moreover, the finite--dimensional reduced dynamics has   
positive maximal Lyapunov exponents $\lambda,\lambda_c$ (the former being 
defined with respect to real time, the latter to the number of collisions). 
These were numerically computed by multiplying matrices obtained from 
linearization of the map (\ref{map}) along a trajectory. 
The exponent $\lambda$ decreases with the number of normal modes taken 
into account, 
because 
bounces become less 
frequent, the two plates going 
to rest, in time average, for $N\to\infty$ (see below). 
On the contrary $\lambda_c$ 
was observed to increase with $N$; indeed, as 
collisions become more distant in time, 
phases change more drastically in between them, 
and their randomization is faster. 
If instead $N$ is
increased keeping the energy per mode $E/N$ fixed, both $\lambda$
and $\lambda_n$ appear to saturate, suggesting that Lyapunov exponents
converge to a finite non zero value in the thermodynamic limit.
The maximal Lyapunov exponents remain positive on reducing $\sigma$,
with 
no stochasticity threshold displayed. However, the time required to
reach a converged value becomes larger, because trajectories need 
more time to fill 
the phase space.

\begin{figure}
\vglue 0cm
\centerline{
\epsfxsize=3.8in
\epsfysize=3.in
\epsffile{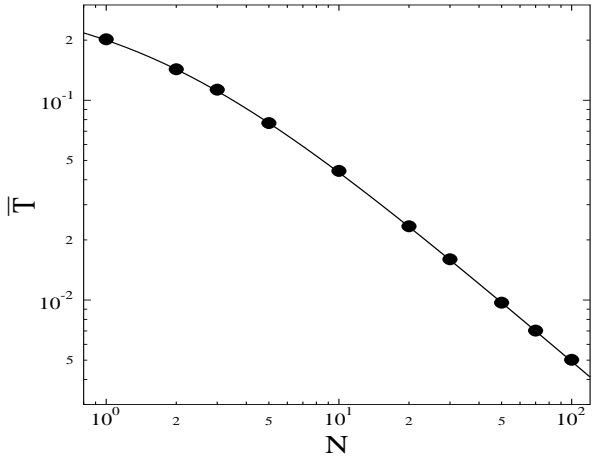}
}
\caption{
Time--average kinetic energy $\overline{T}$ 
of the charged oscillator as a function of the 
number $N$ of odd field modes taken into account. Initial conditions:
$(z,z_1)=(0,0)$, $(\dot {z},\dot {z}{}_1)=(\xi_1,\xi_2)$,
with $\xi_1,\xi_2$ random numbers such that $\frac{1}{2}m(\xi_1^2+
\xi_2^2)=E=1$,  
$(q_n,p_n)=(0,0)$ 
$\forall n=1,2,\ldots,N$; coupling constant 
$\sigma=1$, billiard dimension $R=1$, frequency $\omega_0=5$,
thickness parameter $\delta=0.05$. 
Numerical data follow the equipartition values $\overline{T}=1/(2N+3)$
(full line).
}
\label{fig1}
\end{figure}

Most of our numerical experiments were meant to understand
how the energy is distributed among all the degrees of freedom (in time
average). 
In Fig.\ref{fig1} the time--average kinetic energy 
$\overline{T}=\lim_{t\to\infty}
\,\overline{T}(t)$ of the charged plate is shown as a function 
of the number $N$ of field modes considered. In the above definition, 
$\overline{T}(t)$ is the time average up to time $t$ of 
$\frac{1}{2}m{\dot{z}^2}(t)$, while the 
limit means that the motion has been followed until stabilization of
the time--average. Results are in accordance with the 
equipartition theorem: the total energy is equally shared between the 
$2N+3$ relevant canonical variables. 
For $N\to\infty$ we can extrapolate 
$\overline{T}=0$, that is, the electromagnetic field acts as a friction 
force on the plate. 

\begin{figure}
\vglue 0.cm
\centerline{
\epsfxsize=3.8in
\epsfysize=3.in
\epsffile{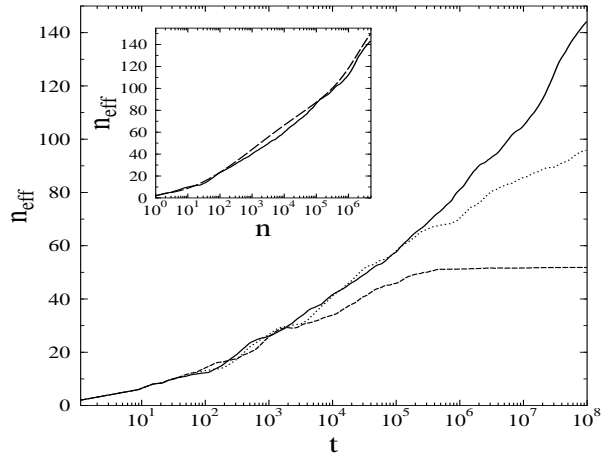}
}
\caption{
Equipartition parameter $n_{eff}$ 
as a function of time,
for $N=50$ (dashed line), $N=100$ (dotted 
line), $N=200$ (full line), with $\sigma=0.3$,
$R=1$, $\omega_0=0.7$, $\delta=0.05$;
initially half of the energy is given to the discharged
plate, half to the first overall normal mode.
The insert shows 
$n_{eff}$ as a function of the number of bounces 
(full line) and its random phase approximation
(dashed line).  
}
\label{fig2}
\end{figure}

The 
approach to equilibrium is not uniform, because the 
relaxation time associated with 
the $n$--th overall normal mode 
increases with $n$. To analyze this increase we have used the equipartition 
indicator
\begin{equation}
n_{eff}(t)=\exp\left\{-\sum_{n=0}^{N+1}\overline{E}_n(t)
\ln\overline{E}_n(t)\right\},
\label{neff}
\end{equation} 
where $\overline{E}_n(t)$ indicates the normalized time
average energy (up to the time $t$) of the $n$--th normal mode
$(\overline{E}_{N+1}$ refers to the energy of the neutral plate).
The parameter $n_{eff}$ is a  measure of  
the number of modes significantly excited at time $t$; if only a finite 
number of modes is considered, it also 
measures the degree of equipartition,  because 
$n_{eff}=1$ if only one normal mode is excited, whereas the maximal value 
$n_{eff}\sim N+3/2$ is only attained in
the presence of complete equipartition. 
As far as the numerical 
simulation is truly representative of the infinite--dimensional system,
$n_{eff}$ appears to increase with $t$ slower than any power, but faster
than logarithmically
(Fig.\ref{fig2}). 

A quantitative description of the relaxation process and 
 analytical estimates 
of expression (\ref{neff}) can be obtained 
from eqn.(\ref{map}) by a random--phase approximation. Let us 
consider an ensemble of trajectories which start at time zero from a single 
normal mode, with randomly distributed phases. Denoting $E_n(\tau)$ 
the ensemble--averaged 
(normalized) energy on the $n-$th normal mode after the $\tau-$th 
collision, and assuming complete randomness of phases at the collision time, 
from (\ref{map}) we get:
\begin{equation}
\label{rf}
E_k(\tau)=
(1-4\vert e_k\vert^2)E_k(\tau-1)+4W(\tau-1)\vert e_k\vert^2,
\end{equation}
where
\begin{equation}
\label{rf1}
e_k=<u_k\vert {\bf e}>;\quad W(\tau)=\sum_k\vert e_k\vert^2 E_k(\tau).
\end{equation}
Recalling the meaning of ${\bf e}$, one easily realizes that  
$W(\tau)$ is proportional to the average 
kinetic energy of the relative motion of the plates.
Eqns. (\ref{rf},\ref{rf1}) can be solved numerically, to find how 
$n_{eff}$ increases with the number of collisions. The result (shown 
by the dashed line in the insert of Fig.(\ref{fig2})) matches quite well with 
the numerical solution of the exact equations of motion, confirming 
the validity of the random phase approximation, hence the chaotic 
nature of dynamics. One can also solve 
(\ref{rf},\ref{rf1}) analytically, by 
implementing a continuous time approximation, plus standard 
Laplace transform techniques. Omitting details, one finds that the 
large--$\tau$ asymptotics  of the solution is determined by the large--$k$ 
asymptotics of the coefficients $a_k$. In  
case of algebraic decay $a_k\sim \vert k\vert^{-\alpha}$, dispensing 
with prefactors which depend on $\epsilon, l, c$
and estimating the average time delay between the $\tau+1$--th 
and the $\tau$-th collision as
$t\sim R/{\sqrt W(\tau)}$, one finds $n_{eff}(t)\sim
t^{\frac{2}{4\alpha+5}}$ \cite{inprep}.
With the charge distribution $f(x)$ used in our numerical simulations, 
we cannot give likewise explicit formulas, 
due to the complicated decay 
of coefficients $a_k$. However, it is possible to prove 
that $n_{eff}$ increases 
with $t$ faster than 
logarithmically, but slower than any power of $t$, as found in Fig.\ref{fig2}.  
Thus the way the relaxation time of modes increases with their frequencies 
is determined by the choice of the charge density.

\begin{figure}
\vglue 0.cm
\centerline{
\epsfxsize=3.8in
\epsfysize=3.in
\epsffile{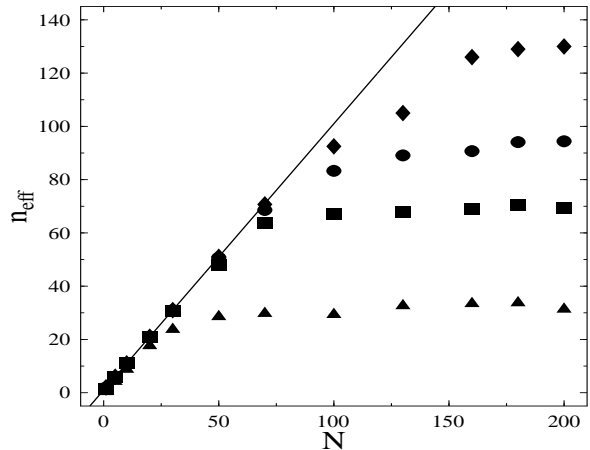}
}
\caption{
Equipartition parameter $n_{eff}$ as a function of the number
$N$ of odd field modes considered, at $t=10^3$ (triangles),
$t=10^5$ (squares), $t=10^7$ (circles), $t=10^8$ (diamonds),
with initial conditions and parameters chosen as in 
Fig.\ref{fig2}.
The full line indicates equipartition values 
$n_{eff}\sim N+3/2$.
}
\label{fig3}
\end{figure}

To verify that the truncated system numerically investigated here really 
represents (up to a certain time) 
the real, infinite--dimensional, model,  
we plotted in Fig.\ref{fig3}  $n_{eff}$ as a function of the 
number $N$ of field oscillators taken into account. We found that, 
at any fixed time,  
$n_{eff}$   
converges on increasing $N$, its limit value giving the number of 
overall normal
modes significantly excited in the infinite--dimensional system.
As this value increases with time, an equilibrium state
is never reached.      

In summary, our 
numerical experiments have exposed a chaotic dynamics in the unusual 
case of an {\it infinite} dimensional, conservative system with a 
{\it finite}  
total energy. The dependence of time-averages on initial conditions
 gets lost, and, for any 
finite--dimensional reduction, the system reaches 
an equilibrium state, with equipartition
of energy among the degrees of freedom of the field and of the matter.

From our results we infer that, in 
the real infinite--dimensional problem, there is a trend towards 
equipartition, the finite energy of matter being removed to higher 
and higher frequencies of the field. However, this process takes place at a 
nonuniform rate, as relaxation times of normal modes increase with 
their frequency. Therefore, as in an old hypothesis of Jeans\cite{JEANS},
a real equilibrium state is never reached.

Artificial though it may appear, our model is actually {\it the simplest, 
one-dim. model} with  charged particles  undergoing elastic collisions inside 
a reflecting enclosure. We believe that more realistic models displaying 
the same basic features will display a similar behaviour.   
We also submit that quantization of this, or similar, field   
models may open a challenging new direction in the field of Quantum Chaos, 
starting from a very old problem.

\end{multicols}

\begin{references}
\bibitem[*]{byline1} Present address: CEA, Service de Physique de
l'Etat Condens\'e, Centre d'Etudes de Saclay, F--91191 Gif--sur--Yvette,
France 
\bibitem{JEANS}J.H.Jeans, {\it The dynamical theory of gases}, 2nd ed., 
Cambridge 1916.
\bibitem{BCL72}
P. Bocchieri, A. Crotti and A. Loinger, Lett. Nuovo Cimento {\bf 4}
(1972) 741.
\bibitem{BLVG74}
 G. Casati, I. Guarneri and F. Valz--Gris, Phys. Rev. A {\bf 16}
(1977) 1237;
G. Benettin and L. Galgani, J. Stat. Phys. {\bf 27} (1982) 153; 
G. Casati, I. Guarneri and F. Valz Gris, J. Stat. Phys. {\bf 30}
(1983) 195; 
R. Livi, M. Pettini, S. Ruffo and A. Vulpiani, J. Phys. A {\bf 20} 
(1987) 577; 
C. Alabiso, M. Casartelli and S. Sello, J. Stat. Phys. {\bf 54}
(1989) 361.
\bibitem{GAFGG89}
L. Galgani, C. Angaroni, L. Forti, A. Giorgilli and F. Guerra,
Phys. Lett. A {\bf 139} (1989) 221;
C. Alabiso, M. Casartelli and A. Scotti, Phys. Lett. A {\bf 147}
(1990) 292. 
\bibitem{P83}
A. Patrascioiu, Phys. Rev. Lett. {\bf 50} (1983) 1879;
K.R.S. Devi and A. Patrascioiu, Physica D {\bf 11} (1984) 359;
A. Patrascioiu, E. Seiler and I.O. Stamatescu, Phys. Rev. A {\bf 31} 
(1985) 1906.
\bibitem{inprep} G. Benenti, G. Casati and I. Guarneri, in preparation.
\end{references}
\end{document}